\documentclass[journal]{IEEEtran}

\usepackage{cite}
\usepackage{amsmath,amsfonts,amssymb}
\usepackage{array}
\usepackage{textcomp}
\usepackage{stfloats}
\usepackage{url}
\usepackage{verbatim}
\usepackage{graphicx}
\usepackage{xcolor}
\usepackage{booktabs}
\usepackage{multirow}
\usepackage{numprint}
\npthousandsep{,}
\npdecimalsign{.}
\usepackage[normalem]{ulem}
\usepackage{siunitx}

\usepackage{caption}
\usepackage{subcaption}

\usepackage[linesnumbered,ruled,vlined]{algorithm2e}
\usepackage{algpseudocode}
\usepackage{algorithmicx}
\usepackage{algcompatible}

\def\tsc#1{\csdef{#1}{\textsc{\lowercase{#1}}\xspace}}
\tsc{WGM}
\tsc{QE}
\tsc{EP}
\tsc{PMS}
\tsc{BEC}
\tsc{DE}

\begin{document}

\title{CHAOS: Controlled Hardware fAult injectOr System for gem5}

\author{
    Elio Vinciguerra, Enrico Russo, Giuseppe Ascia, Maurizio Palesi, \IEEEmembership{Senior Member, IEEE}
    \thanks{Elio Vinciguerra, Enrico Russo, Giuseppe Ascia, and Maurizio Palesi are with the Department of Electrical, Electronic and Computer Engineering, University of Catania, Catania, Italy. (email: elio.vinciguerra@phd.unict.it, enrico.russo@unict.it, giuseppe.ascia@unict.it, maurizio.palesi@unict.it)}
}

\maketitle

\begin{abstract}
Fault injectors are essential tools for evaluating the reliability and resilience of computing systems. They enable the simulation of hardware and software faults to analyze system behavior under error conditions and assess its ability to operate correctly despite disruptions. Such analysis is critical for identifying vulnerabilities and improving system robustness. \textit{CHAOS} is a modular, open-source, and fully configurable fault injection framework designed for the gem5 simulator. It facilitates precise and systematic fault injection across multiple architectural levels, supporting comprehensive evaluations of fault tolerance mechanisms and resilience strategies. Its high configurability and seamless integration with gem5 allow researchers to explore a wide range of fault models and complex scenarios, making \textit{CHAOS} a valuable tool for advancing research in dependable and high-performance computing systems.
\end{abstract}


\IEEEpubidadjcol

\section{Introduction}
\label{sec:Introduction}

Fault tolerance has become a fundamental requirement in the design of modern computing systems, especially in domains where reliability, safety, and availability are non-negotiable, such as aerospace~\cite{aerospace}, automotive~\cite{automotive}, healthcare~\cite{healthcare}, and cloud infrastructures~\cite{cloud}. As systems grow in complexity and scale, they become increasingly susceptible to faults caused by hardware defects, transient errors (e.g., radiation-induced soft errors), and software bugs~\cite{emb_sys}. The ability to evaluate how a system reacts to such faults is therefore essential to ensure its robustness and correctness.

Fault injectors (FIs) are specialized tools that simulate hardware or software faults during the execution of a system. By artificially introducing errors, such as bit flips in memory, corrupted architectural registers, or erroneous instructions, FIs allow engineers to observe fault propagation, identify potential vulnerabilities, and validate the effectiveness of error detection and recovery mechanisms. This kind of testing is especially important when traditional verification methods fall short, as in the case of rare or non-deterministic faults that are difficult to reproduce.

In this context, simulation-based fault injection has gained prominence due to its flexibility and low cost. Unlike physical fault injection, which requires expensive hardware setups, simulation allows precise control over fault parameters and system observability. Among the available simulators, gem5~\cite{gem5} stands out as a modular and highly configurable platform that supports a wide range of processor architectures and system configurations. Its extensive use in computer architecture research has made it a \emph{de facto} standard for evaluating design trade-offs and system behavior under various workloads.

Given its popularity, several fault injection frameworks have been developed for gem5~\cite{FIMSIM, GeFIN, GemFI, gem5Approx, MARVEL}. However, many of these are now obsolete, lack support for recent gem5 versions, are limited in functionality, or are closed-source, making them difficult to adopt and extend. This fragmentation presents a significant barrier to researchers and practitioners who need reliable, flexible, open, and up-to-date FI tools.

To address these challenges, we present Controlled Hardware fAult injectOr System (\textit{CHAOS})~\cite{chaos2025}: an open-source, modular, and extensible fault injection framework for gem5. \textit{CHAOS} is designed to support all Instruction Set Architectures (ISAs) available in gem5, and to provide fine-grained control over injection timing, fault types, and target components.

The key contributions of this paper are as follows:
\begin{itemize}
\item We introduce \textit{CHAOS}, a novel fault injection framework for gem5 that is open-source, easily configurable, and compatible with multiple ISAs.
\item We provide a detailed description of its architecture, including the mechanisms for injecting faults in architectural registers, memory, and cache subsystems.
\item We evaluate \textit{CHAOS} through extensive simulations and performance evaluation, demonstrating its low overhead and flexibility in diverse use cases.
\item We release \textit{CHAOS} to the community to encourage reproducibility, future development, and collaborative fault injection research.
\end{itemize}

The remainder of this paper is structured as follows: Sec.~\ref{sec:background} presents the necessary background and related work, Sec.~\ref{sec:design} details the design and implementation of \textit{CHAOS}, Sec.~\ref{sec:evaluation} discusses the evaluation of the proposed framework, and finally, Sec.~\ref{sec:conclusions} outlines the conclusions.

\section{Background and Related Work}
\label{sec:background}

\subsection{Fault Injectors}
\label{subsec:FIbackground}

A FI is a crucial tool for assessing the reliability and resilience of computing systems. Its primary purpose is to simulate hardware and software faults to observe how the system responds and evaluate its ability to function correctly despite errors. This type of analysis is essential for identifying vulnerabilities and enhancing system robustness.

Faults can be categorized into several types~\cite{faultsType}:
\begin{itemize}
    \item \textit{Transient faults}: Short-lived errors induced by external disturbances such as radiation or electrical noise, which can momentarily alter the correct operation of hardware components, affecting data, control signals, or logic states. \emph{Example:} A cosmic ray strike may flip a bit in memory, causing a temporary error that is corrected upon the next write.
    \item \textit{Intermittent faults}: Errors that occur sporadically over time, making them difficult to predict and diagnose. \emph{Example:} A loose connection in a solder joint may cause occasional data corruption during specific temperature or voltage conditions.
    \item \textit{Permanent faults}: Hardware failures that cause lasting damage to a component. \emph{Example:} A burnt-out transistor in a logic gate causes a permanent logic error in computations.
\end{itemize}

Regardless of the fault type or its granularity -- whether it affects a single bit or multiple bits -- the observable effects on system behavior can be categorized into the following outcomes:

\begin{itemize}
    \item \textit{Crash}: The error is severe enough to cause a complete system shutdown.
    \item \textit{Detected Unrecoverable Error} (DUE): The fault is detected, but the system lacks the ability to recover, leading to failure.
    \item \textit{Silent Data Corruption} (SDC): The fault leads to undetected data corruption, resulting in incorrect outputs without any warning.
    \item \textit{Masked}: The fault has no impact on program execution, and the final result is identical to that of a fault-free run.
    \item \textit{Timeout}: The system fails to complete execution within the expected time.
\end{itemize}

A comprehensive understanding of fault types and their associated outcomes is crucial for the effective design of fault injection campaigns. It allows for a systematic and targeted exploration of system behavior under a wide range of fault conditions, enabling the identification of weaknesses and the validation of error-handling mechanisms. Moreover, fault injection plays a pivotal role not only in the validation phase but also in shaping the architecture of next-generation computing systems, fostering the development of designs that exhibit higher fault tolerance, robustness, and overall dependability.

\subsection{gem5}

gem5~\cite{gem5} is a widely adopted open-source microarchitectural simulator known for its flexibility and detailed modeling of processor architectures and ISAs. It enables instruction- and cycle-accurate analysis of processor behavior, accurately representing elements such as pipelines, caches, and branch predictors. Its highly modular and configurable design makes it ideal for architectural exploration and system-level research.

Supporting multiple ISAs, including ARM, MIPS, POWER, RISC-V, SPARC, and X86, gem5 uses a discrete-event simulation model that allows researchers to evaluate various systems and workloads under controlled conditions. The framework’s extensible architecture and object-oriented structure enable straightforward customization and rapid prototyping.

At the core of this modularity are SimObjects, representing both hardware components (e.g., cores, caches, interconnects) and abstract entities (e.g., workloads, process contexts). Each SimObject is defined by a Python class, used for configuration and instantiation, and a C++ class implementing its internal logic. A common interface for initialization, statistics, and checkpointing streamlines integration and code reuse across the platform.

\subsection{Fault Injectors in gem5}

Several fault injection (FI) tools have been built on top of gem5, targeting different reliability aspects and fault models.

FIMSIM~\cite{FIMSIM} enables microarchitectural-level analysis across various fault models but is based on the legacy M5 simulator; not compatible with modern gem5 versions. GeFIN~\cite{GeFIN} supports massive campaigns across transient, permanent, and intermittent faults in multiple hardware structures. Useful for comprehensive reliability studies in Gem5; supports detailed classification of fault outcomes. GemFI~\cite{GemFI} provides extensible fault injection in functional and cycle-accurate modes with fast checkpointing, but is based on older gem5 releases and lacks advanced fault effect analysis. gem5-Approxilyzer~\cite{gem5Approx} focuses on efficient transient bit-level error analysis with output quality classification, yet cannot handle concurrent faults or special register injections. gem5-MARVEL~\cite{MARVEL} offers consolidated fault injection for heterogeneous SoC architectures supporting all major ISAs and vulnerability analysis, but remains closed-source.

Only GemFI and gem5-Approxilyzer are open-source, but both are outdated and lack modularity, limiting compatibility with recent gem5 releases. FIMSIM, GeFIN, and gem5-MARVEL are closed-source, which fundamentally restricts their adaptability, prevents community-driven extensions, and hinders broader adoption across the research landscape.

Overall, no existing framework offers the combined benefits of open-source accessibility, modern gem5 compatibility (20+), modularity, and comprehensive fault injection support. \textit{CHAOS} addresses these limitations through a fully open-source and modular infrastructure compatible with all gem5 ISAs, enabling fine-grained, multi-level fault injection across architectural registers, caches, and main memory. This makes \textit{CHAOS} a flexible, extensible, and accessible solution for reliability research.

\section{Design and Implementation Strategy}
\label{sec:design}

This section introduces \textit{CHAOS}, emphasizing the strengths of the proposed system and the fault injection techniques it employs. \textit{CHAOS} is composed of three modules. The first, \textit{CHAOReg}, disrupts application execution by injecting faults into various architectural registers at runtime. The second module, \textit{CHAOSCache}, injects faults into cache subsystem during execution. Finally, \textit{CHAOSMem} handles fault injection into memory during execution. The following subsections provide a detailed discussion of each individual module and the specific mechanisms they use to perform fault injection.

\subsection{CHAOReg}
\label{subsec:Reg}

This module is designed to inject faults into the CPU architectural registers during execution, randomly altering their contents by applying a fault mask. To provide a fundamental set of register-level modifications for simulating various error scenarios during execution, three distinct types of faults are currently supported: The first type, a \textit{bit flip}, causes one or more bits in the register to change their value, either flipping from 0 to 1 or from 1 to 0. This type of alteration simulates errors that may occur due to environmental factors, such as radiation-induced soft errors. The second type is referred to as \textit{stuck at zero}. In this case, one or more bits within the register are permanently forced to 0, which can lead to incorrect or missing data depending on the affected bits’ role in the system. Similarly, the third fault type, \textit{stuck at one}, forces one or more bits to remain at 1, causing potential data corruption or unexpected behavior when the system attempts to use those bits.

The integration of \textit{CHAOReg} into the gem5 simulator is achieved by following the pseudocode outlined in Algorithm~\ref{alg:CHAOSReg}, which defines its behavior. One of the main input parameters is \textit{probability}, a floating-point value between 0 and 1, sets the likelihood of activating \textit{CHAOS} in a given clock cycle. To further control when faults are injected, \textit{start} and \textit{end} define the range of clock cycles during which \textit{CHAOS} is allowed to operate. The \textit{fault\_type} parameter specifies the type of fault to be injected, while \textit{mask} represents the bitmask applied to the target register. If set to \textit{0}, a random bitmask is generated instead. In such cases, \textit{faulty\_bits} defines the number of bits affected by the randomly generated mask. To refine targeting, \textit{target\_class} specifies the class of architectural registers that can be targeted. Finally, \textit{$PC_{target}$} defines the program counter (PC) address at which \textit{CHAOS} should be activated, ensuring precise fault injection at specific execution points.

The \textit{CHAOReg} module is triggered at random clock cycles, sampled according to the \textit{probability} parameter. Once a fault event is activated, the module checks whether the current clock cycle falls within the user-defined range or whether the PC matches the \textit{$PC_{target}$} parameter; if either condition holds, the fault injection procedure is initiated (line~\ref{line:active}). The module then determines the architectural register class in which to inject the fault (line~\ref{line:class}) and randomly selects the target register to corrupt (line~\ref{line:randomReg}). Subsequently, the fault mask is evaluated and, if it is set to zero, a new mask is randomly generated (line~\ref{line:mask}), together with the selection of the fault type to be injected (line~\ref{line:type}). Once all parameters have been resolved, the fault is injected into the selected architectural register: depending on the chosen fault type, the injector applies either a bit flip (line~\ref{line:bitflip}), a stuck-at-zero fault (line~\ref{line:stuckat0}), or a stuck-at-one fault (line~\ref{line:stuckat1}). It is important to emphasize that, while a bit flip models a transient fault, stuck-at-zero and stuck-at-one faults are permanent. To preserve the persistent effect of these permanent faults, they are stored in a dedicated data structure that is periodically inspected to reapply the fault (line~\ref{line:markpermanent1} and~\ref{line:markpermanent2}). Finally, the next fault injection event is scheduled, with the inter-injection interval determined by the configured probability.

\begin{algorithm}[t]
  \caption{Integration of \textit{CHAOReg} into gem5.}
\label{alg:CHAOSReg}
\KwData{probability, start, end, fault\_type, mask, faulty\_bits, target\_class, $PC_{target}$}
  \DontPrintSemicolon
    \If{$\bigl(PC_{\text{target}} = 0 \textbf{ and } \text{start} \le \text{curCycle()} \le \text{end}\bigr)$ \textbf{ or } $\bigl(PC(t) = PC_{\text{target}}\bigr)$}{
    \label{line:active}
      \If{$target\_class = \text{random}$}{
      \label{line:class}
        $target\_class \gets \text{Sample}\{\text{integer}, \text{floating-point}\}$\;
      }
      
      $r \gets$ random register in $target\_class$\;
      \label{line:randomReg}
      
      \If{$mask = 0$}{
        $mask \gets \text{random\_mask}(\text{faulty\_bits})$\;
        \label{line:mask}
      }
      \If{$fault\_type = \text{random}$}{
        $fault\_type \gets \text{Sample}\{\text{bit\_flip}, \text{stuck\_at\_0}, \text{stuck\_at\_1}\}$\;
        \label{line:type}
      }
      \uIf{$fault\_type = \text{bit\_flip}$}{
        $\text{Value}(r) \gets \text{Value}(r) \oplus mask$\;
        \label{line:bitflip}
      }
      \uElseIf{$fault\_type = \text{stuck\_at\_0}$}{
        $\text{Value}(r) \gets \text{Value}(r) \wedge \neg mask$\;
        \label{line:stuckat0}
        \text{MarkPermanentFault}($r, mask, \text{stuck\_at\_0}$)\;
        \label{line:markpermanent1}
      }
      \ElseIf{$fault\_type = \text{stuck\_at\_1}$}{
      \label{line:stuckat1}
        $\text{Value}(r) \gets \text{Value}(r) \vee mask$\;
        \label{line:markpermanent2}
        \text{MarkPermanentFault}($r, mask, \text{stuck\_at\_1}$)\;
      }
    }

    schedule next inject fault after random delay proportional to probability\;
\end{algorithm}

\subsection{CHAOSCache}
\label{subsec:Cache}
The \textit{CHAOSCache} module is designed to inject faults at the cache memory level, enabling the simulation of errors that can affect the performance and reliability of the caching mechanism. Similar to \textit{CHAOReg}, it supports the same fault types, allowing for a consistent approach to fault injection across different components of the system. This includes simulating bit flips, stuck-at-zero, and stuck-at-one faults, which can help assess the system’s behavior under various failure conditions. The integration of \textit{CHAOSCache} into the gem5 simulator follows the pseudocode outlined in Algorithm~\ref{alg:CHAOSCache}, which defines its behavior and execution flow.

To enable fine-grained control over the fault injection process, \textit{CHAOSCache} exposes a set of configuration parameters that determine how faults are applied to the cache memory. These parameters include \textit{cache}, \textit{probability}, \textit{start}, \textit{end}, \textit{fault\_type}, \textit{corruption\_size}, \textit{mask}, and \textit{faulty\_bits}. Their semantics closely mirror those of \textit{CHAOReg}, as described in the previous section, with two notable exceptions: \textit{cache}, which specifies the target cache instance in which faults are injected, and \textit{corruption\_size}, which defines the number of bytes to be corrupted. By appropriately tuning these parameters, users can precisely control when and how faults are introduced, thereby enabling flexible experimentation and systematic analysis of cache-memory vulnerabilities.

The \textit{CHAOSCache} module is triggered at pseudo-random clock cycles drawn according to the \textit{probability} parameter. When a fault injection event is issued, the module first verifies that the current clock cycle lies within the user-specified interval. If this condition holds, a valid block is randomly selected in the target cache (line~\ref{line:cache_block}). For the selected block, the subsequent stages are repeated for as many bytes as specified by the configuration, corresponding to the number of bytes to be corrupted. At each iteration, a byte offset within the block is chosen (line~\ref{line:cache_byte}), and, analogously to \textit{CHAOReg}, the mask (line~\ref{line:cache_mask}) and the fault type (line~\ref{line:cache_type}) are resolved; if either parameter is left at its default value, it is instantiated randomly. The fault is then injected according to the selected type, namely as a bit flip (line~\ref{line:cache_bitflip}), a stuck-at-zero fault (line~\ref{line:cache_stuckat0}), or a stuck-at-one fault (line~\ref{line:cache_stuckat1}). As in the register-level case, stuck-at faults model permanent errors, and each injected stuck-at fault is therefore recorded in a dedicated data structure that is periodically scanned to reapply the corresponding modification (line~\ref{line:cache_markpermanent1} and~\ref{line:cache_markpermanent2}). Finally, the next injection event is scheduled, with the inter-injection interval determined by the configured probability parameter.

\begin{algorithm}[t]
      \caption{Integration of \textit{CHAOSCache} into gem5.}
  \label{alg:CHAOSCache}
  \KwData{cache, probability, start, end, fault\_type, corruption\_size, mask, faulty\_bits}
  \DontPrintSemicolon

  \If{$(\text{start} \le \text{curCycle()} \le \text{end}\bigr)$}{
    $B \gets \text{SampleValidBlock}($cache$)$\;
    \label{line:cache_block}

    \For{$i \gets 1$ \KwTo corruption\_size}{
      $b \gets \text{SampleRandomByte}(B)$\;
      \label{line:cache_byte}
      \If{$mask = 0$}{
        $mask \gets \text{random\_mask}(\text{faulty\_bits})$\;
        \label{line:cache_mask}
      }
      \If{$fault\_type = \text{random}$}{
        $fault\_type \gets \text{Sample}\{\text{bit\_flip}, \text{stuck\_at\_0}, \text{stuck\_at\_1}\}$\;
        \label{line:cache_type}
      }

      \uIf{$fault\_type = \text{bit\_flip}$}{
        $\text{Value}(b) \gets \text{Value}(b) \oplus mask$\;
        \label{line:cache_bitflip}
      }
      \uElseIf{$fault\_type = \text{stuck\_at\_0}$}{
        $\text{Value}(b) \gets \text{Value}(b) \wedge \neg mask$\;
        \label{line:cache_stuckat0}
        \text{MarkPermanentFault}($B, b, mask, \text{stuck\_at\_0}$)\;
        \label{line:cache_markpermanent1}
      }
      \ElseIf{$fault\_type = \text{stuck\_at\_1}$}{
        $\text{Value}(b) \gets \text{Value}(b) \vee mask$\;
        \label{line:cache_stuckat1}
        \text{MarkPermanentFault}($B, b, mask, \text{stuck\_at\_1}$)\;
        \label{line:cache_markpermanent2}
      }
    }
  }

  schedule next cache fault injection after random delay proportional to probability\;
\end{algorithm}

\subsection{CHAOSMem}
\label{subsec:memory}

The \textit{CHAOSMem} module is designed to inject faults at the main memory level, enabling the simulation and analysis of errors that can occur within the memory subsystem. This module facilitates the exploration of memory-related failures, which can have a significant impact on system performance and reliability. Similar to \textit{CHAOReg} and \textit{CHAOSCache}, it supports three distinct fault types: \textit{bit flip}, \textit{stuck at zero}, and \textit{stuck at one}. These faults replicate common failure scenarios typically encountered in real-world hardware systems, helping to assess the system’s robustness and ability to handle potential memory errors.

As in the previously described components, \textit{CHAOSMem} relies on a set of configuration parameters to define its behavior. These parameters -- \textit{probability}, \textit{start}, \textit{end}, \textit{target\_start}, \textit{target\_end}, \textit{fault\_type}, \textit{mask}, and \textit{faulty\_bits} -- have semantics that mirror those introduced earlier, with the exception of \textit{target\_start} and \textit{target\_end}, which denote the bounds of the address range within which faults may be injected.

The integration of \textit{CHAOSMem} into the gem5 simulator follows a process similar to the previews one. However, there are key differences in how memory faults are injected, which are outlined in Algorithm~\ref{alg:CHAOSMem}.
The \textit{CHAOSMem} module is triggered at pseudo-random clock cycles drawn according to the \textit{probability} parameter. When a fault injection event is issued, the module first verifies that the current clock cycle lies within the user-specified interval. If this condition is met, a random address is selected within the configured target range (line~\ref{line:mem_addr}), and the corresponding data byte is read from memory (line~\ref{line:mem_read}). The mask (line~\ref{line:mem_mask}) and the fault type (line~\ref{line:mem_type}) are then examined; if either parameter remains at its default value, it is instantiated randomly. The fault is subsequently injected according to the selected model, namely as a bit flip (line~\ref{line:mem_bitflip}), a stuck-at-zero fault (line~\ref{line:mem_stuckat0}), or a stuck-at-one fault (line~\ref{line:mem_stuckat1}). As in the previous cases, stuck-at faults represent permanent errors and are therefore recorded in a dedicated data structure that is periodically traversed to reapply the corresponding modification when necessary (line~\ref{line:mem_markpermanent1} and~\ref{line:mem_markpermanent2}). Finally, the modified data is written back to memory (line~\ref{line:mem_write}), and the next fault injection event is scheduled, with its occurrence determined by the configured probability.

\begin{algorithm}[t]
    \caption{Integration of \textit{CHAOSMem} into gem5.}
  \label{alg:CHAOSMem}
  \KwData{probability, start, end, target\_start, target\_end, fault\_type, mask, faulty\_bits}
  \DontPrintSemicolon

  \If{$(\text{start} \le \text{curCycle()} \le \text{end}\bigr)$}{
    $addr \gets \text{SampleRandomAddr}(target\_start, target\_end)$\;
    \label{line:mem_addr}
    
    $data \gets$ ReadByte($addr$)\;
    \label{line:mem_read}
    
    \If{$mask = 0$}{
      $mask \gets \text{random\_mask}(faulty\_bits)$\;
      \label{line:mem_mask}
    }
    \If{$fault\_type = \text{random}$}{
      $fault\_type \gets \text{Sample}\{\text{bit\_flip}, \text{stuck\_at\_0}, \text{stuck\_at\_1}\}$\;
      \label{line:mem_type}
    }

    \uIf{$fault\_type = \text{bit\_flip}$}{
      $data \gets data \oplus mask$\;
      \label{line:mem_bitflip}
    }
    \uElseIf{$fault\_type = \text{stuck\_at\_0}$}{
      $data \gets data \wedge \neg mask$\;
      \label{line:mem_stuckat0}
      \text{MarkPermanentFault}($addr, mask, \text{stuck\_at\_0}$)\;
      \label{line:mem_markpermanent1}
    }
    \ElseIf{$fault\_type = \text{stuck\_at\_1}$}{
      $data \gets data \vee mask$\;
      \label{line:mem_stuckat1}
      \text{MarkPermanentFault}($addr, mask, \text{stuck\_at\_1}$)\;
      \label{line:mem_markpermanent2}
    }
    WriteByte($addr$, $data$)\;
    \label{line:mem_write}
  }

  schedule next memory fault injection after random delay proportional to probability\;
\end{algorithm}

\section{Experimental Evaluation}
\label{sec:evaluation}
In this section, we present the results of several simulations conducted using \textit{CHAOS} in gem5. We analyze metrics such as the overhead on total simulation time and the impact of injected faults on the simulation, considering variations in \textit{CHAOS} parameters like injection probability.

\subsection{Experimental Setup}
\label{subsec:setup}

The following section provides an overview of the gem5 setup used and the benchmarks selected for testing.

The simulated system features a \SI{1}{\giga\hertz} clock frequency, DDR3 DRAM memory of \SI{512}{\mebi\byte}, and an Out-of-Order (O3) CPU. The cache hierarchy includes an L1 Instruction (L1I) cache of \SI{16}{\kibi\byte}, an L1 Data (L1D) cache of \SI{64}{\kibi\byte}, and an L2 cache of \SI{256}{\kibi\byte}. This setup is modeled using the default parameters provided by gem5.

While \textit{CHAOS} demonstrates compatibility with any ISA supported by gem5, the subsequent analysis focuses exclusively on experiments conducted using the RISC-V~\cite{RISC-VSpec1, RISC-VSpec2} architecture.

The workloads selected for the gem5 simulation come from the MiBench suite~\cite{MiBench}, and include eight benchmarks, implemented in C: Bitcount, Blowfish, Dijkstra, JPEG, Patricia, Qsort, SHA and Susan.

These benchmarks were chosen to represent a diverse set of application behaviors, enabling comprehensive evaluation of system reliability under various fault conditions. To support this analysis, Table~\ref{tab:metrics} summarizes key performance metrics for each benchmark, offering a detailed characterization of their computational profiles and resource utilization patterns.

\begin{table*}[]
\caption{Performance Metrics of Critical Algorithms for Fault Tolerance Testing}
\centering
\begin{tabular}{@{}ccccccccc@{}}
\toprule
Metric                                      & Bitcount & Blowfish & Dijkstra & JPEG  & Patricia & Qsort & SHA   & Susan \\ \midrule
Total CPU clock cycles                      & 196M     & 125M     & 258M     & 364M  & 260M     & 153M  & 125M  & 180M  \\
Total executed instructions                 & 519M     & 311M     & 460M     & 540M  & 218M     & 300M  & 406M  & 376M  \\
Average cycles per instruction              & 0.38     & 0.40     & 0.56     & 0.67  & 1.19     & 0.51  & 0.31  & 0.48  \\
Branch misprediction ratio {[}\%{]}         & 1.87     & 0.74     & 0.1      & 11.53 & 6.92     & 1.65  & 3.59  & 11.68 \\
Committed memory reads {[}\%{]}             & 5.21     & 22.32    & 23.72    & 22.6  & 17.82    & 17.81 & 11.53 & 26.65 \\
Committed memory writes {[}\%{]}            & 0.002    & 11.75    & 11.56    & 9.28  & 11.24    & 13.31 & 4.49  & 4.68  \\
Committed integer ALU instructions {[}\%{]} & 94.79    & 65.93    & 64.71    & 67.01 & 69.12    & 66.35 & 83.98 & 65.75 \\
L1D cache miss rate {[}\%{]}                & 0.03     & 0.02     & 5        & 1.12  & 0.23     & 2.53  & 0.01  & 0.4   \\
L1I cache miss rate {[}\%{]}                & 0.003    & 0.005    & 0.002    & 0.006 & 23.53    & 0.003 & 0.002 & 0.002 \\
L2 cache miss rate {[}\%{]}                 & 77.18    & 51.8     & 70.52    & 75.14 & 0.59     & 70.59 & 79.96 & 98.23 \\ \bottomrule
\end{tabular}
    \label{tab:metrics}
\end{table*}

The eight benchmarks exhibit markedly different microarchitectural behaviors in terms of control flow, ALU utilization, and pressure on the memory hierarchy, which makes this set well suited for a comparative architectural study. The reported metrics span a spectrum from highly compute‑bound kernels with excellent cache locality to workloads dominated by irregular memory access and complex control flow.

Bitcount and SHA represent the class of strongly compute‑bound workloads: almost all committed instructions are integer ALU operations (94.79\% for Bitcount and 83.98\% for SHA), and the average cycles per instruction (CPI) are very low (0.38 and 0.31, respectively). The negligible L1D miss rates (around 0.03\% and 0.01\%) indicate excellent data locality, while the high L2 miss rates (77.18\% and 79.96\%) suggest working sets that exceed the L2 capacity yet still exhibit intense reuse within L1.

Blowfish shares many characteristics of cryptographic kernels, combining a high fraction of ALU instructions (65.93\%) with a moderately low CPI (0.40), which is consistent with a regular, well‑pipelineable computational flow. Its extremely low L1D miss rate (0.02\%) indicates highly regular data access patterns, whereas the still high L2 miss rate (51.8\%) implies that reuse is confined to relatively small blocks that frequently oscillate between L1 and deeper levels of the memory hierarchy.

Dijkstra, Patricia, and Qsort are representative of graph and data‑structure algorithms with complex control flow, as evidenced by significant fractions of memory reads (17–24\%) and writes (11–13\%) relative to total committed instructions. Dijkstra exhibits an intermediate CPI (0.56) but a high L1D miss rate (5\%) and a very high L2 miss rate (70.52\%), which is consistent with the irregular access patterns typical of graph exploration. Patricia, while maintaining a comparable level of ALU utilization (69.12\%), stands out for having the worst CPI in the entire set (1.19) and an extremely high L1I miss rate (23.53\%) despite an almost ideal L2 miss rate (0.59\%), suggesting an instruction footprint that exceeds I‑cache capacity but fits entirely within L2. Qsort shows an intermediate profile, with CPI 0.51, a moderate branch misprediction ratio (1.65\%), and a non‑negligible L1D miss rate (2.53\%), while its high L2 miss rate (70.59\%) reflects the scattered access patterns induced by recursive partitioning.

JPEG and Susan constitute clearly data‑intensive workloads, with some of the highest percentages of memory reads (22–27\%) and very high branch misprediction ratios (11.53\% for JPEG and 11.68\% for Susan), indicating complex, data‑dependent control flow. JPEG is the most resource‑demanding benchmark in the suite, with 364M clock cycles, 540M committed instructions, and a CPI of 0.67, coupled with an above‑average L1D miss rate (1.12\%), which is consistent with the less local memory access patterns typical of image compression. Susan, in turn, combines a high branch misprediction ratio with an almost maximal L2 miss rate (98.23\%) despite a relatively low L1D miss rate (0.4\%), making it a particularly demanding workload for the deeper levels of the memory hierarchy and for the branch prediction subsystem.

Overall, the benchmark suite covers the full spectrum from almost purely computational kernels with excellent locality (Bitcount, SHA), through cryptographic workloads that exert strong pressure on L2 (Blowfish, SHA), to data‑structure algorithms with highly irregular access patterns (Dijkstra, Patricia, Qsort) and multimedia or vision applications with high memory and control intensity (JPEG, Susan). This heterogeneity makes the suite especially suitable for reliability and fault‑tolerance studies, as it enables observation of system behavior under differentiated stress on the frontend, the computational backend, and the memory hierarchy during fault‑injection campaigns or exploration of alternative architectural configurations.

\subsection{CHAOS Overhead}
The integration of \textit{CHAOS} into the simulation workflow introduces a runtime overhead, arising from the execution of additional instrumentation code, as previously described. This section provides a detailed quantitative characterization of this overhead.

Simulations were conducted for 1 million clock cycles, and for each run the time spent on fault injection was measured and recorded. It should be noted that the observed overhead is not continuously proportional to the configured fault injection probability. Instead, additional time is incurred only when a fault is actually injected, \emph{i.e.}, when the probability condition is satisfied during simulation. Consequently, the reported overhead corresponds exclusively to scenarios in which fault injection occurred and reflects the cost of executing the associated fault-handling routines, including access to control structures and data manipulation.
 
To evaluate the worst-case conditions, fault masks were generated at runtime, with both the number of faulty bits and the fault types randomized.
The average overhead per injected fault respect to the faulty free simulation was measured to be 0.0004\% for \textit{CHAOReg}, 0.0008\% for \textit{CHAOSCache}, and 0.0004\% for \textit{CHAOSMem}. Moreover, when a permanent fault is injected, a dedicated routine is invoked to monitor the fault at clock-cycle granularity, ensuring that its effect is maintained throughout the execution. The overhead associated with this monitoring, for each injected permanent fault, is $6.6\times 10^{-6}\%$ for \textit{CHAOReg}, $6.1 \times 10^{-6}\%$ for \textit{CHAOSCache}, and $8.1\times 10^{-6}\%$ for \textit{CHAOSMem}.

These low overhead values demonstrate the efficiency of \textit{CHAOS} in performing precise and controlled fault injections with minimal impact on simulation performance. This limited disruption helps to ensure that resilience evaluations remain realistic and computationally tractable, thereby supporting scalable and practically relevant reliability studies.

\subsection{Using CHAOS}
In this section, we examine the utilization of the \textit{CHAOS} module, specifically focusing on its fault injection mechanism within the gem5-simulated RISC-V processor, where target is a single bit randomly selected at runtime.
To ensure reliable results, a fault campaign was conducted for each \textit{CHAOS} module individually, isolating the effects of \textit{CHAOReg}, \textit{CHAOSMem}, and \textit{CHAOSCache} on L1D, L1I, and L2 caches. 

To facilitate comprehensive fault injection campaigns, three distinct injection tiers were established based on the methodology in~\cite{SFI}: \textit{Low}, targeting a margin of error $e = 5\%$ with a 95\% confidence interval; \textit{Medium}, targeting $e = 5\%$ with a 99\% confidence interval; and \textit{High}, targeting $e = 1\%$ with a 99\% confidence interval. These parameters necessitate sample sizes of 384, 663, and 16,587 faults, respectively.

Subsequently, the impact of the \textit{CHAOS} module on the simulation outcomes was analyzed, considering the effects discussed in Sec.~\ref{subsec:FIbackground}.

In the following discussion, no distinction will be made between crash and DUE, as both lead to an irreversible execution halt. The term crash will be used to collectively refer to both cases. Furthermore, throughout this fault campaign, no simulations resulted in an outcome classified as a timeout; therefore, this category will not be considered.

\subsubsection{Crashes}

\begin{figure*}[htbp]
    \centering
    \begin{minipage}[t]{0.90\textwidth}
        \centering
        \includegraphics[width=\textwidth]{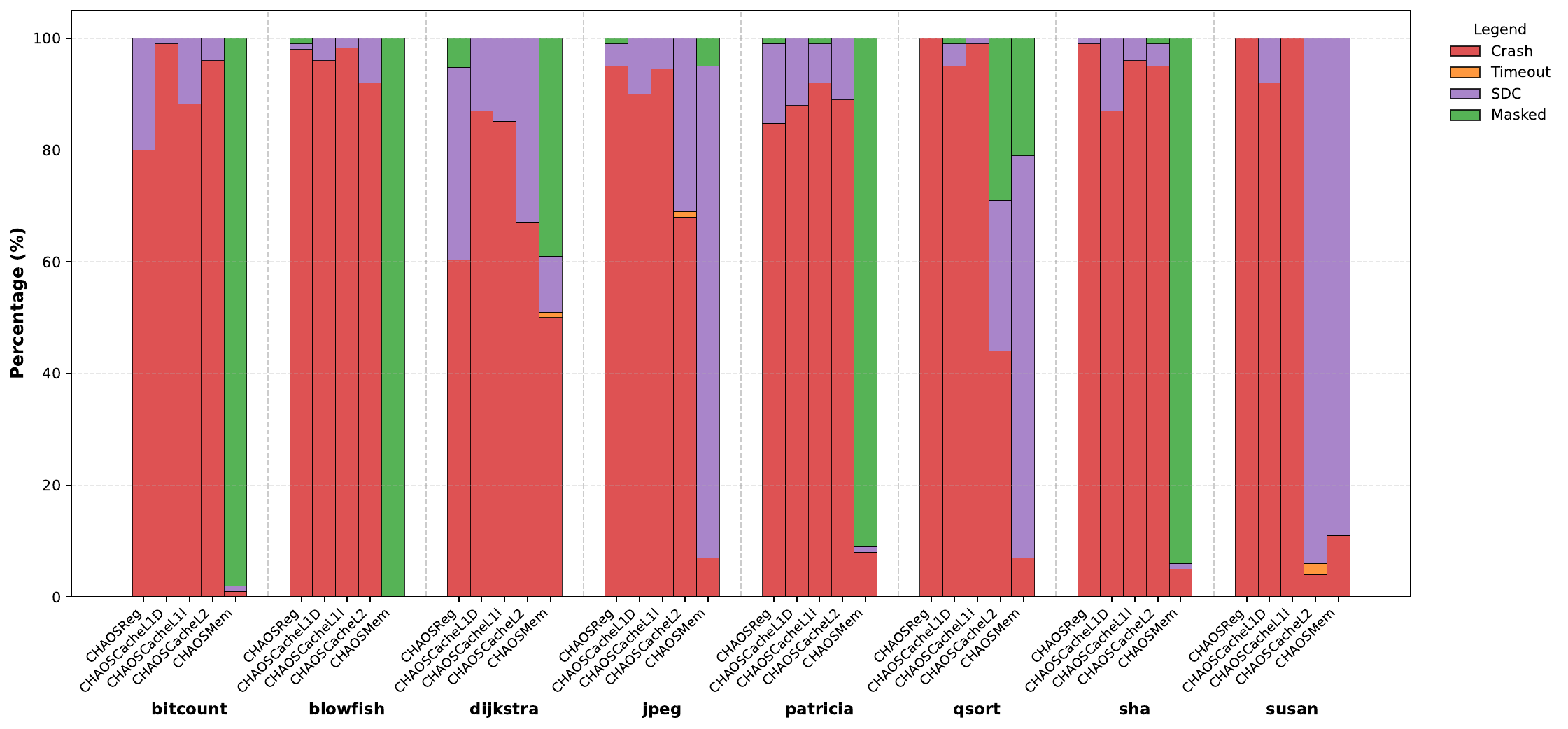}
        \caption{Fault Classification by Injector and Application: High Fault Injection Probability.}
        \label{fig:faultClassHigh}
    \end{minipage}
    \hfill
    \begin{minipage}[t]{0.90\textwidth}
        \centering
        \includegraphics[width=\textwidth]{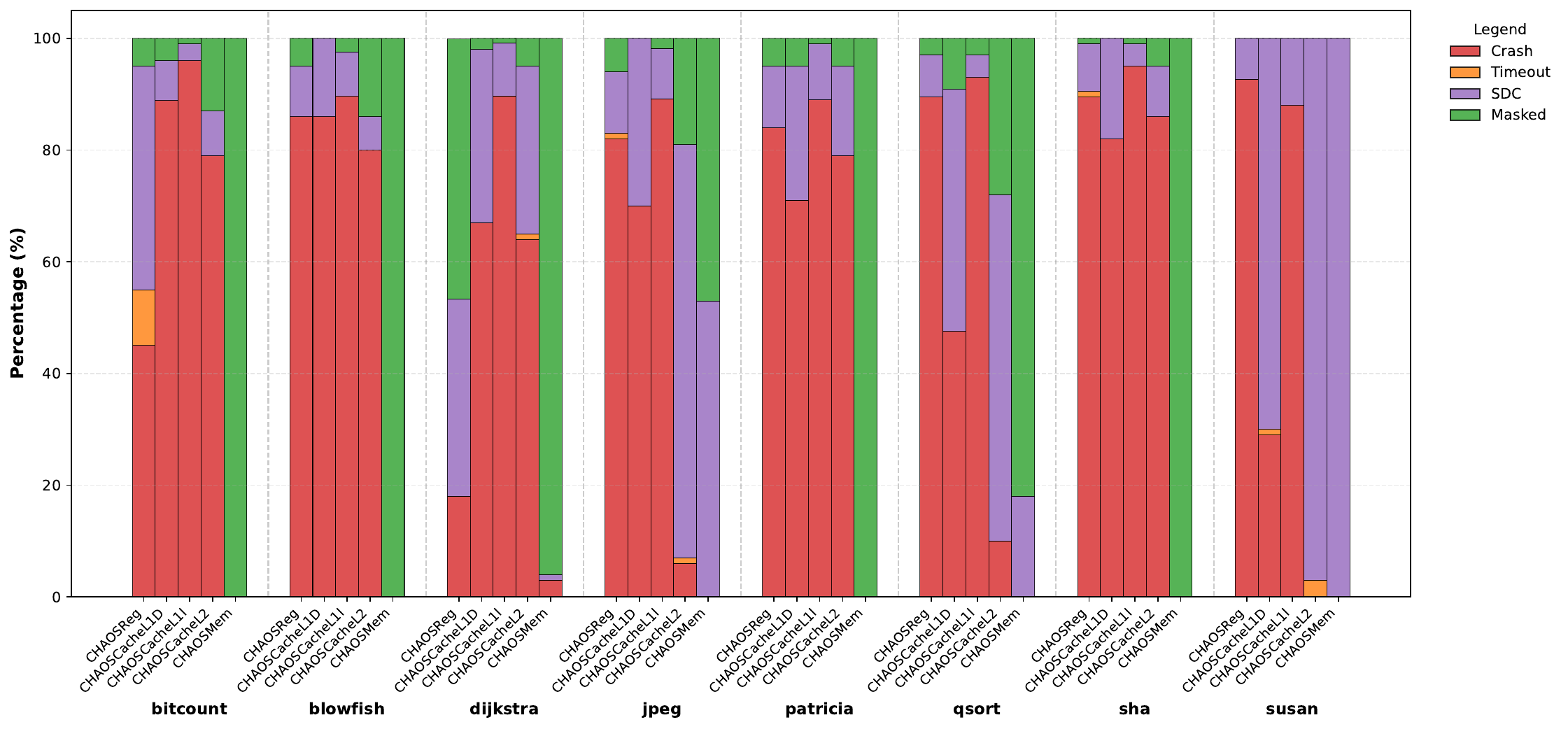}
        \caption{Fault Classification by Injector and Application: Medium Fault Injection Probability.}
        \label{fig:faultClassMedium}
    \end{minipage}
    \hfill
    \begin{minipage}[t]{0.90\textwidth}
        \centering
        \includegraphics[width=\textwidth]{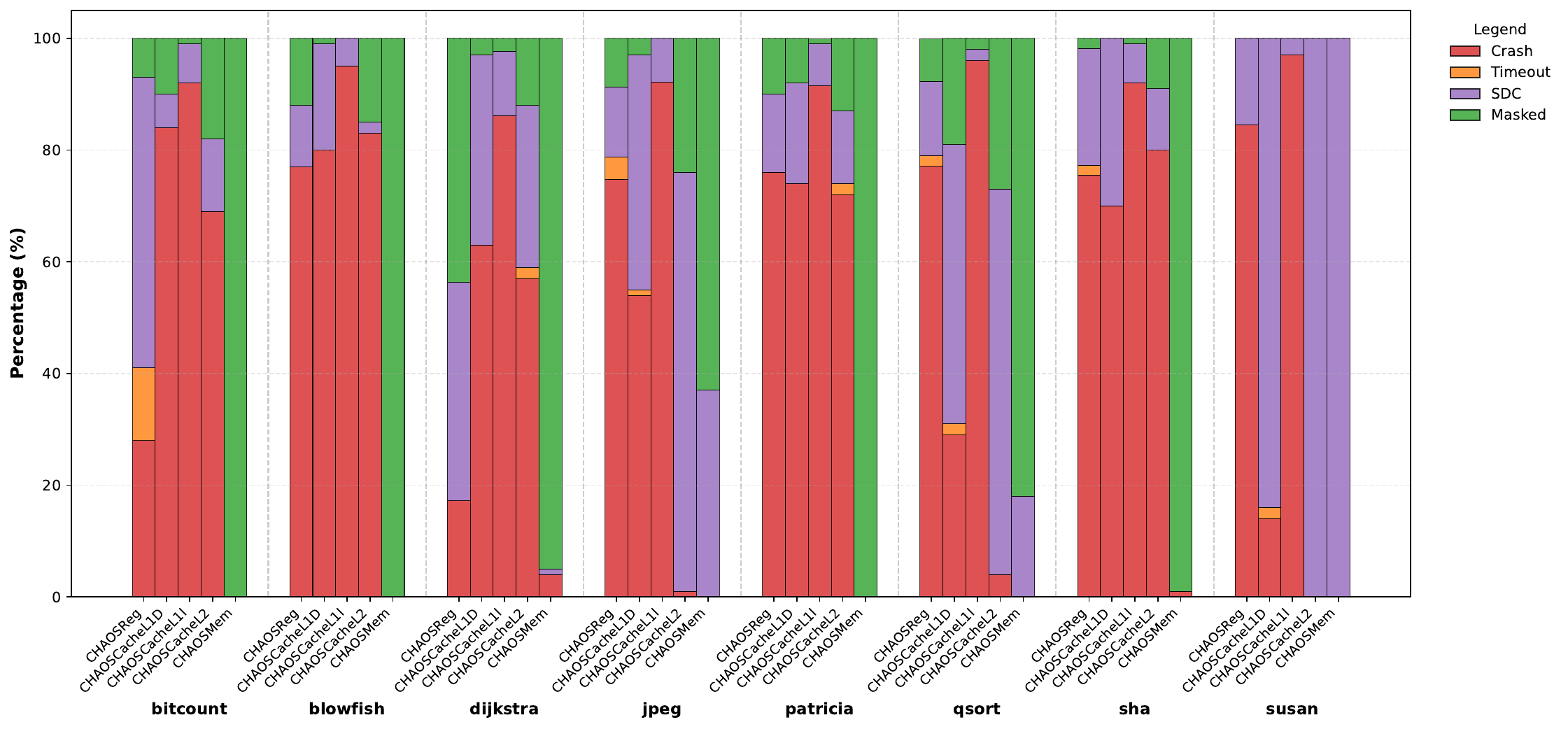}
        \caption{Fault Classification by Injector and Application: Low Fault Injection Probability.}
        \label{fig:faultClassLow}
    \end{minipage}
\end{figure*}

We begin by analyzing the occurrence of crashes. The likelihood of such events varies significantly depending on the target component and the fault injection probability. Figures~\ref{fig:faultClassHigh}, \ref{fig:faultClassMedium}, and \ref{fig:faultClassLow} present the breakdown of fault outcomes for High, Medium, and Low injection probabilities, respectively.

Architectural registers demonstrate a direct correlation between injection probability and crash rates. At High probabilities (Figure~\ref{fig:faultClassHigh}), the "Crash" outcome (represented in red) is dominant across most applications. This is particularly evident in \textit{jpeg}, \textit{qsort}, and \textit{bitcount}, where crashes account for the overwhelming majority of faults. This confirms that frequent corruption of registers -- which may hold Program Counters or critical pointers -- leads to immediate termination.
However, behavior diverges at Medium (Figure~\ref{fig:faultClassMedium}) and Low probabilities (Figure~\ref{fig:faultClassLow}). While some applications, such as \textit{jpeg}, maintain a notable crash rate, others like \textit{dijkstra} and \textit{susan} show a significant reduction. At low probabilities, the system is more likely to mask the fault or produce an SDC rather than abort. This indicates that sparse register faults are less likely to corrupt a critical "live" value that triggers an exception.

The L1I cache exhibits the highest sensitivity to faults. As shown across all three figures, the crash rate for \textit{CHAOSCacheL1I} remains consistently high regardless of injection probability. This indicates that corruption in the instruction cache is almost invariably fatal. Irrespective of fault frequency, altering an instruction opcode or operand typically results in an illegal operation or a segmentation fault, causing immediate application failure.

Faults in data caches (L1D and L2) also yield a very high percentage of crashes, confirming the critical nature of these components. At High probabilities, the behavior is nearly identical to that of the L1I cache, with crash outcomes dominating results. However, a subtle trend emerges at Low probabilities (Figure~\ref{fig:faultClassLow}). While crashes remain the primary outcome, their prevalence decreases slightly in favor of SDC (purple) and Masked (green) outcomes compared to the High probability scenario. This suggests that while heavy corruption of data caches guarantees a crash, sporadic faults have a small probability of affecting non-critical data or data that does not cause an immediate exception, although the risk of a crash remains predominant.

In stark contrast to registers and caches, Main Memory exhibits a very low occurrence of crashes across all scenarios. In Figures~\ref{fig:faultClassHigh}, \ref{fig:faultClassMedium}, and \ref{fig:faultClassLow}, the crash rate for \textit{CHAOSMem} is negligible. Even at High probabilities, due to the vast size of the main memory, faults are statistically less likely to hit a segment currently vital for program flow control (such as a return address stack or a critical pointer). Instead, faults here tend to result in SDCs or remain masked (unused data), rather than causing the immediate termination observed with cache or register faults.

\begin{table}[t]
\caption{HPCs monitored and recorded within the statistics log file provided by gem5.}
\centering
\begin{tabular}{@{}cc@{}}
\toprule
Name          & Description                                       \\ \midrule
mcycle        & cycles number                                    \\
mtime         & time                                             \\
minstret      & instructions retired                             \\
mhpmcounter4  & integer load instruction retired                 \\
mhpmcounter5  & integer store instruction retired                \\
mhpmcounter7  & system instruction retired                       \\
mhpmcounter8  & integer arithmetic instruction retired           \\
mhpmcounter9  & conditional branch retired                       \\
mhpmcounter10 & JAL instruction retired                          \\
mhpmcounter11 & JALR instruction retired                         \\
mhpmcounter12 & integer multiplication instruction retired       \\
mhpmcounter13 & integer division instruction retired             \\
mhpmcounter14 & floating point load \& store instruction retired \\
mhpmcounter15 & floating point other instruction retired         \\
mhpmcounter22 & branch/jump target mispredicion                  \\
mhpmcounter27 & instruction cache miss                           \\
mhpmcounter28 & data cache miss or memory-mapped I/O access      \\
mhpmcounter29 & data cache writeback                             \\
mhpmcounter30 & instruction TLB miss                             \\
mhpmcounter31 & data TLB miss                                    \\ \bottomrule
\end{tabular}
    \label{tab:hpmgem5}
\end{table}

\begin{table*}[]
    \centering
    \small
    
    \caption{Analysis of HPC Variations: \textit{CHAOSReg}}
    \label{tab:chaosreg_sdc_masked}
    \resizebox{\linewidth}{!}{
    \begin{tabular}{@{}ccccccccccccccccc@{}}
    \toprule
    \textbf{} & \multicolumn{2}{c}{\textbf{Bitcount}} & \multicolumn{2}{c}{\textbf{Blowfish}} & \multicolumn{2}{c}{\textbf{Dijkstra}} & \multicolumn{2}{c}{\textbf{JPEG}} & \multicolumn{2}{c}{\textbf{Patricia}} & \multicolumn{2}{c}{\textbf{Qsort}} & \multicolumn{2}{c}{\textbf{SHA}} & \multicolumn{2}{c}{\textbf{Susan}} \\
    \midrule
     & SDC & Masked & SDC & Masked & SDC & Masked & SDC & Masked & SDC & Masked & SDC & Masked & SDC & Masked & SDC & Masked \\
    \textbf{High} & 1.32 & - & 0.86 & 0.86 & 4.44 & 0.00 & 2.73 & 0.30 & 0.57 & 0.00 & 0.01 & 0.00 & 0.00 & - & 0.41 & - \\
    \textbf{Medium} & 0.82 & 0.00 & 0.89 & 0.86 & 5.65 & 0.00 & 0.40 & 0.30 & 0.04 & 0.00 & - & - & 0.00 & 0.00 & - & - \\
    \textbf{Low} & 1.27 & 0.00 & 0.81 & 0.86 & 8.05 & 0.00 & 0.32 & 0.30 & 0.00 & 0.00 & 26.01 & 0.00 & 0.02 & 0.00 & 0.44 & - \\
    \bottomrule
    \end{tabular}
    }
    
    \vspace{0.2cm}
    
    \caption{Analysis of HPC Variations: \textit{CHAOSCacheL1D}}
    \label{tab:chaoscachel1d_sdc_masked}
    \resizebox{\linewidth}{!}{
    \begin{tabular}{@{}ccccccccccccccccc@{}}
    \toprule
    \textbf{} & \multicolumn{2}{c}{\textbf{Bitcount}} & \multicolumn{2}{c}{\textbf{Blowfish}} & \multicolumn{2}{c}{\textbf{Dijkstra}} & \multicolumn{2}{c}{\textbf{JPEG}} & \multicolumn{2}{c}{\textbf{Patricia}} & \multicolumn{2}{c}{\textbf{Qsort}} & \multicolumn{2}{c}{\textbf{SHA}} & \multicolumn{2}{c}{\textbf{Susan}} \\
    \midrule
     & SDC & Masked & SDC & Masked & SDC & Masked & SDC & Masked & SDC & Masked & SDC & Masked & SDC & Masked & SDC & Masked \\
    \textbf{High} & 0.14 & - & 0.96 & - & 0.01 & - & 0.32 & - & 0.01 & - & 20841.92 & 0.00 & 0.00 & - & 0.07 & - \\
    \textbf{Medium} & 1.63 & 0.01 & 0.70 & - & 0.12 & 0.00 & 0.32 & - & 0.08 & 0.00 & 83211.58 & 0.00 & 0.01 & - & 0.08 & - \\
    \textbf{Low} & 2.10 & 0.01 & 0.77 & 0.86 & 0.01 & 0.00 & 0.32 & 0.30 & 0.07 & 0.00 & 52867.90 & 0.00 & 0.05 & - & 0.06 & - \\
    \bottomrule
    \end{tabular}
    }
    
    \vspace{0.2cm}
    
    \caption{Analysis of HPC Variations: \textit{CHAOSCacheL1I}}
    \label{tab:chaoscachel1i_sdc_masked}
    \resizebox{\linewidth}{!}{
    \begin{tabular}{@{}ccccccccccccccccc@{}}
    \toprule
    \textbf{} & \multicolumn{2}{c}{\textbf{Bitcount}} & \multicolumn{2}{c}{\textbf{Blowfish}} & \multicolumn{2}{c}{\textbf{Dijkstra}} & \multicolumn{2}{c}{\textbf{JPEG}} & \multicolumn{2}{c}{\textbf{Patricia}} & \multicolumn{2}{c}{\textbf{Qsort}} & \multicolumn{2}{c}{\textbf{SHA}} & \multicolumn{2}{c}{\textbf{Susan}} \\
    \midrule
     & SDC & Masked & SDC & Masked & SDC & Masked & SDC & Masked & SDC & Masked & SDC & Masked & SDC & Masked & SDC & Masked \\
    \textbf{High} & 7767.44 & - & 0.81 & - & 0.25 & - & 4.94 & - & 1.87 & 0.00 & 0.00 & - & 0.00 & - & 0.98 & - \\
    \textbf{Medium} & 0.00 & 0.00 & 4.39 & 0.86 & 0.10 & 0.00 & 2.27 & 0.30 & 1.36 & 0.00 & 0.01 & 0.00 & 71.77 & 0.00 & - & - \\
    \textbf{Low} & 6.69 & 0.00 & 2.04 & - & 2460.71 & 0.00 & 7.81 & - & 10.63 & 0.00 & 0.03 & 0.00 & 0.03 & 0.00 & 0.23 & - \\
    \bottomrule
    \end{tabular}
    }
    
    \vspace{0.2cm}
    
    \caption{Analysis of HPC Variations: \textit{CHAOSCacheL2}}
    \label{tab:chaoscachel2_sdc_masked}
    \resizebox{\linewidth}{!}{
    \begin{tabular}{@{}ccccccccccccccccc@{}}
    \toprule
    \textbf{} & \multicolumn{2}{c}{\textbf{Bitcount}} & \multicolumn{2}{c}{\textbf{Blowfish}} & \multicolumn{2}{c}{\textbf{Dijkstra}} & \multicolumn{2}{c}{\textbf{JPEG}} & \multicolumn{2}{c}{\textbf{Patricia}} & \multicolumn{2}{c}{\textbf{Qsort}} & \multicolumn{2}{c}{\textbf{SHA}} & \multicolumn{2}{c}{\textbf{Susan}} \\
    \midrule
     & SDC & Masked & SDC & Masked & SDC & Masked & SDC & Masked & SDC & Masked & SDC & Masked & SDC & Masked & SDC & Masked \\
    \textbf{High} & 5285.90 & - & 4.09 & - & 0.21 & - & 2.62 & - & 0.33 & - & 148.99 & 0.00 & 0.06 & 0.00 & 0.06 & - \\
    \textbf{Medium} & 0.07 & 0.01 & 0.59 & 0.76 & 1.22 & 0.00 & 0.31 & 0.30 & 5.50 & 0.00 & 905.74 & 0.00 & 0.28 & 0.01 & 0.05 & - \\
    \textbf{Low} & 988.34 & 0.01 & 0.79 & 0.75 & 0.08 & 0.00 & 0.32 & 0.30 & 19.21 & 0.00 & 253.98 & 0.00 & 0.11 & 0.01 & 0.05 & - \\
    \bottomrule
    \end{tabular}
    }
    
    \vspace{0.2cm}
    
    \caption{Analysis of HPC Variations: \textit{CHAOSMem}}
    \label{tab:chaosmem_sdc_masked}
    \resizebox{\linewidth}{!}{
    \begin{tabular}{@{}ccccccccccccccccc@{}}
    \toprule
    \textbf{} & \multicolumn{2}{c}{\textbf{Bitcount}} & \multicolumn{2}{c}{\textbf{Blowfish}} & \multicolumn{2}{c}{\textbf{Dijkstra}} & \multicolumn{2}{c}{\textbf{JPEG}} & \multicolumn{2}{c}{\textbf{Patricia}} & \multicolumn{2}{c}{\textbf{Qsort}} & \multicolumn{2}{c}{\textbf{SHA}} & \multicolumn{2}{c}{\textbf{Susan}} \\
    \midrule
     & SDC & Masked & SDC & Masked & SDC & Masked & SDC & Masked & SDC & Masked & SDC & Masked & SDC & Masked & SDC & Masked \\
    \textbf{High} & 0.00 & 0.00 & - & 0.86 & 1.86 & 0.00 & 0.32 & 0.30 & 0.00 & 0.00 & 6.58 & 0.00 & 0.03 & 0.00 & 0.05 & - \\
    \textbf{Medium} & - & 0.00 & - & 0.86 & 0.00 & 0.00 & 0.31 & 0.30 & - & 0.00 & 0.00 & 0.00 & - & 0.00 & 0.05 & - \\
    \textbf{Low} & - & 0.00 & - & 0.86 & 0.01 & 0.00 & 0.31 & 0.30 & - & 0.00 & 1.89 & 0.00 & - & 0.00 & 0.05 & - \\
    \bottomrule
    \end{tabular}
    }
\end{table*}

\subsubsection{Timeouts}

Following the analysis of crashes, we address Timeouts, which represent simulations that failed to complete within the predefined execution time limit. This outcome typically arises when a fault corrupts loop termination conditions, branching logic, or wait-states, causing the application to enter an infinite loop or a deadlock.

Data analysis reveals that timeouts are a minority outcome compared to crashes and SDCs, yet they exhibit a distinct pattern. They are most prevalent in the \textit{Bitcount} application, accounting for 13\% of cases at Low probability and 10\% at Medium probability. Minor occurrences (ranging from 1\% to 4\%) are also observed in \textit{JPEG}, \textit{Qsort}, and \textit{SHA}.

A key observation is the inverse correlation between fault injection probability and timeout frequency. Timeouts are exclusively observed at Low and Medium probabilities and disappear entirely at High probabilities across all benchmarks. This suggests that sparse faults have a higher chance of subtly altering control flow (e.g., flipping a loop condition) without invalidating the instruction stream. Conversely, at High probabilities, the corruption is sufficiently severe and widespread that it almost invariably triggers an immediate hardware exception (e.g., Illegal Instruction or Segmentation Fault), causing the system to crash before it can hang indefinitely.

\subsubsection{SDC and Masked outcomes}

Faults leading to SDC are particularly insidious as they do not trigger crashes or noticeable system failures. The program continues execution appearing nominally correct, yet silently produces corrupted results. This characteristic makes SDCs difficult to detect and potentially critical in safety-critical applications.

During the fault injection campaign, each observed outcome was systematically compared against a reference golden run obtained from gem5 executed without \textit{CHAOS} modules. This comparison enabled the identification of fault-induced deviations that, while not resulting in a crash or timeout, yielded incorrect final outputs.

However, the injection of a fault does not invariably degrade the system state. \textit{CHAOReg} may target architectural registers that are effectively dead (unused or overwritten before read). Similarly, \textit{CHAOSCache} and \textit{CHAOSMem} may corrupt memory lines that are never accessed or are subsequently overwritten. In these instances, the fault is completely masked, and the simulation retains a consistent state.

\subsubsection{Exploiting HPC Counters}

To analyze the internal impact of \textit{CHAOS} on simulations classified as SDC or Masked, we leverage the Hardware Performance Counters (HPC)~\cite{RISC-V2, Perf} registers available in the RISC-V ISA. Even when faults are masked or do not prevent program termination, they often induce noticeable perturbations in micro-architectural activity. This correlation suggests that HPC registers can serve as effective proxies for fault detection (system observability), acting as sentinels for anomalies even when the final output appears uncompromised.

We quantify the impact of faults by calculating the mean absolute percentage variation of HPCs. This metric represents the magnitude of the deviation between the fault-free ($h_{\text{ff}}$) and faulty ($h_{\text{f}}$) executions across the $n=20$ monitored counters (listed in Tab.~\ref{tab:hpmgem5}). The variation is computed as follows:

\begin{equation} \Delta_{\text{mean}} = \frac{1}{n} \sum_{i=1}^{n} \left| \frac{h^{(i)}{\text{ff}} - h^{(i)}{\text{f}}}{h^{(i)}_{\text{ff}}} \right| \cdot 100 \label{eq:variation} \end{equation}

By using the absolute value, this metric captures the extent of the divergence regardless of whether the faulty run exhibited higher or lower activity than the golden run. A value of zero indicates identical behavior, while higher values signify severe disruptions in the execution flow. The results are detailed in Tables~\ref{tab:chaosreg_sdc_masked} through \ref{tab:chaosmem_sdc_masked}.

\paragraph{Analysis of Registers} Table~\ref{tab:chaosreg_sdc_masked} presents the variations induced by \textit{CHAOSReg}. Generally, architectural register faults result in moderate HPC variations, as the corruption is often localized to data values rather than control flow. However, \textit{Qsort} presents a notable exception at Low probabilities, exhibiting a significant variation of 26.01\%. This suggests that sparse register faults in this benchmark may corrupt critical loop counters or indices, altering the execution path without causing a crash. \textit{Dijkstra} also shows a distinct trend, with variations increasing from 4.44\% at High probability to 8.05\% at Low probability. In contrast, benchmarks such as \textit{Patricia} and \textit{SHA} show negligible variations (near 0\%), indicating a high degree of execution stability despite register corruption.

\paragraph{Analysis of Caches} The impact of cache faults is markedly more severe and erratic compared to registers, particularly for the L1 Data Cache (Tab.~\ref{tab:chaoscachel1d_sdc_masked}). Here, \textit{Qsort} stands out as an extreme outlier, with variations exceeding 20,000\% in the High scenario and reaching a staggering 83,211\% at Medium probabilities. This magnitude of error suggests that faults in the data cache likely corrupt sorting pointers or array boundaries, causing the algorithm to execute vastly more instructions, potentially entering near-infinite loops that eventually terminate, resulting in correct execution flow (no crash) but massive performance deviations. Other benchmarks in L1D remain relatively stable, with variations generally below 2\%.

In the L1 Instruction Cache (Tab.~\ref{tab:chaoscachel1i_sdc_masked}), \textit{Bitcount} and \textit{Dijkstra} exhibit significant sensitivity. \textit{Bitcount} shows a massive variation at High probabilities (7,767\%), while \textit{Dijkstra} spikes at Low probabilities (2,460\%). This confirms that corruption in the instruction cache can redirect the program counter to valid but unintended code paths, drastically inflating instruction counts. \textit{SHA} also displays a specific sensitivity at Medium probabilities (71.77\%).

The L2 Cache (Tab.~\ref{tab:chaoscachel2_sdc_masked}) mirrors these patterns but with attenuated magnitudes for some workloads. \textit{Bitcount} again shows high sensitivity at High probabilities (5,285\%) and Low probabilities (988\%). \textit{Qsort} maintains significant variations across all probabilities (ranging from 148\% to 905\%), confirming the systemic vulnerability of these workloads to cache disturbances.

\paragraph{Analysis of Main Memory} In stark contrast to the volatile behavior observed in caches, the Main Memory (\textit{CHAOSMem}, Tab.~\ref{tab:chaosmem_sdc_masked}) demonstrates high resilience. Variations across almost all benchmarks are negligible (mostly 0.00\% to 0.30\%). The only minor exception is \textit{Qsort}, which shows a variation of 6.58\% at High probability. This reinforces the observation that while main memory faults may corrupt data values leading to SDCs, they rarely alter the program's control flow or execution duration sufficient to skew HPCs significantly.

Collectively, these results highlight a crucial insight: silent faults, even when non-fatal, can induce massive "under-the-hood" disruptions. While a program like \textit{Qsort} or \textit{Bitcount} might not crash, the internal execution path can be wildly divergent from the golden run. This validates the use of HPC registers as a powerful, non-intrusive mechanism for detecting SDCs that output validation alone might miss or detect too late.

\subsection{Impact of Random Multi-Bit Injection}
\label{sec:random_bit_analysis}

Having analyzed the effects of single-bit faults, we extend our evaluation to a scenario involving random multi-bit injection. In this configuration, the number of bits flipped per injection is selected stochastically. This significantly increases the entropy of the corruption, testing the system's resilience against more aggressive data degradation.

\subsubsection{Fault Outcome Distribution}

\begin{figure*}[htbp]
    \centering
    \begin{minipage}[t]{0.85\textwidth}
        \centering
        \includegraphics[width=\textwidth]{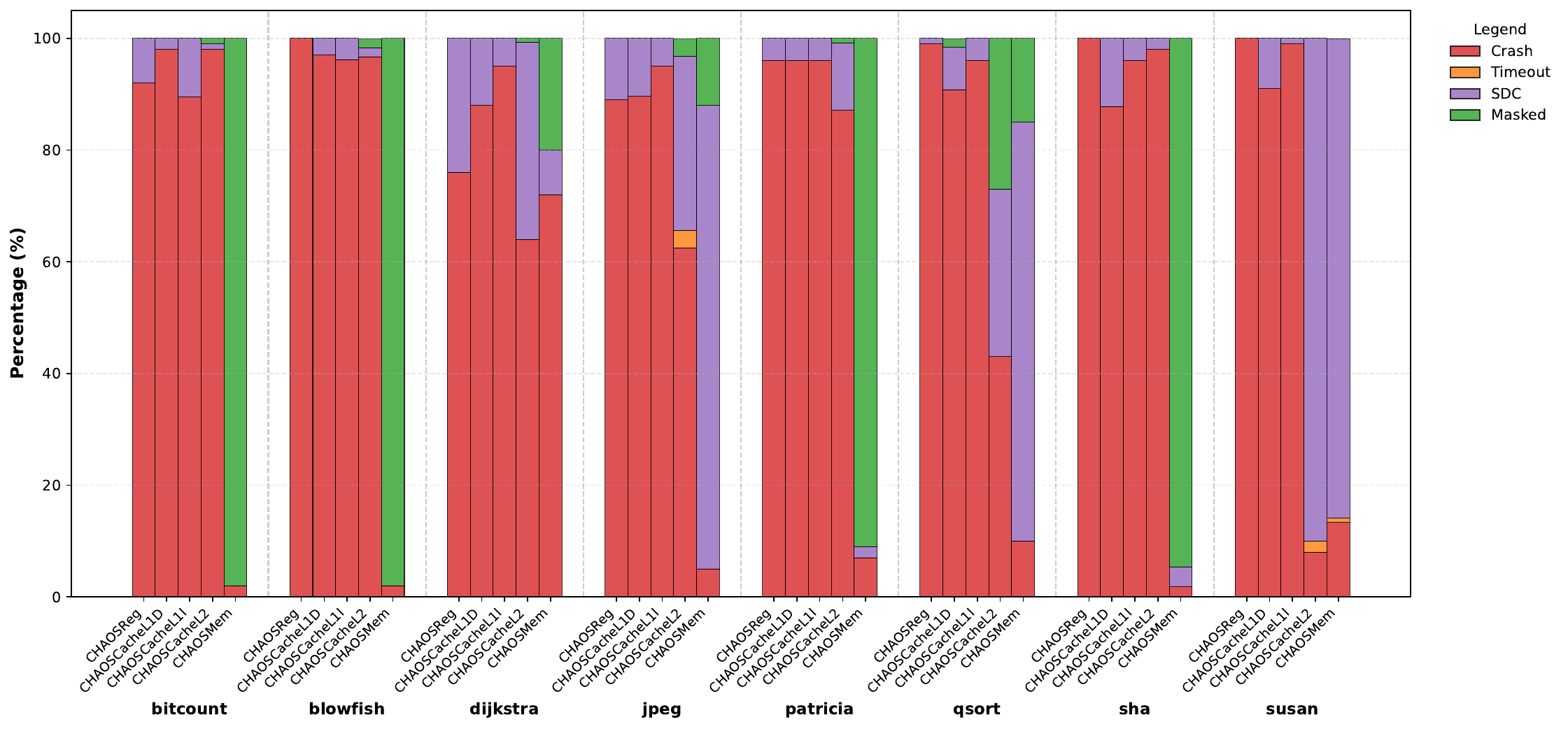}
        \caption{Fault Classification by Injector and Application under Random Multi-Bit Injection: High Fault Injection Probability.}
        \label{fig:faultClassHigh_multi}
    \end{minipage}
    \hfill
    \begin{minipage}[t]{0.85\textwidth}
        \centering
        \includegraphics[width=\textwidth]{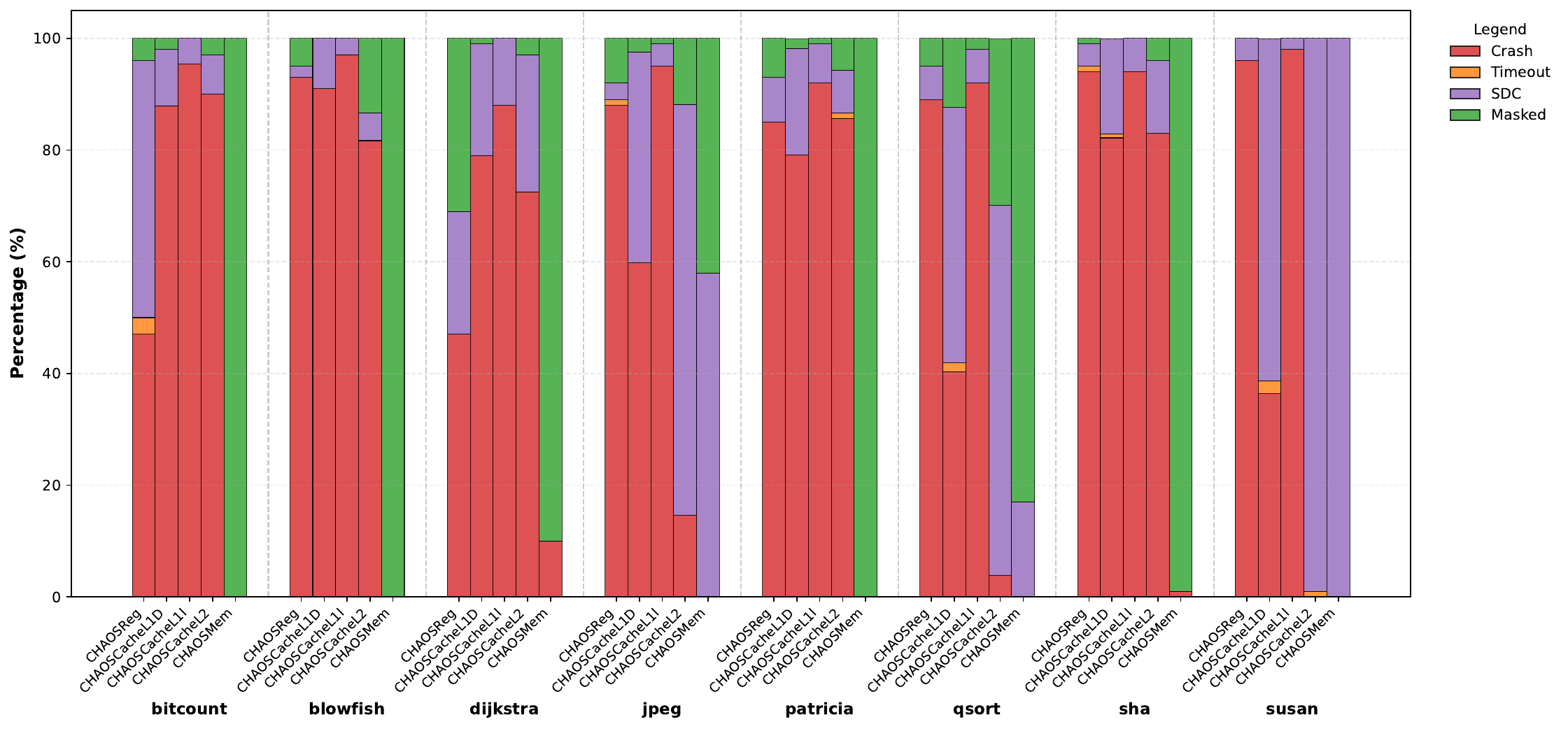}
        \caption{Fault Classification by Injector and Application under Random Multi-Bit Injection: Medium Fault Injection Probability.}
        \label{fig:faultClassMedium_multi}
    \end{minipage}
    \hfill
    \begin{minipage}[t]{0.85\textwidth}
        \centering
        \includegraphics[width=\textwidth]{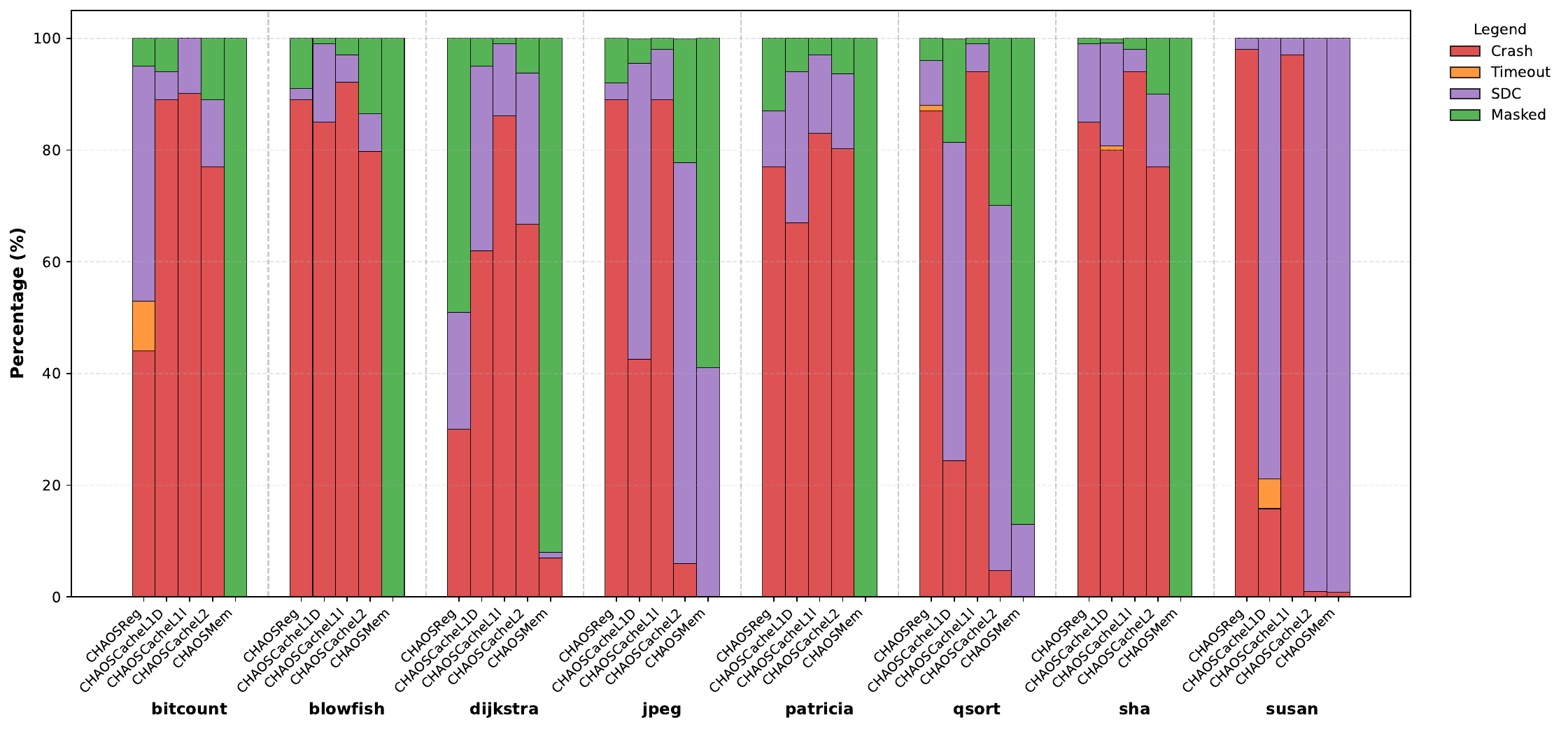}
        \caption{Fault Classification by Injector and Application under Random Multi-Bit Injection: Low Fault Injection Probability.}
        \label{fig:faultClassLow_multi}
    \end{minipage}
\end{figure*}

The transition to multi-bit faults exacerbates the instability of the system, though with component-specific nuances. Figures~\ref{fig:faultClassHigh_multi}, \ref{fig:faultClassMedium_multi}, and \ref{fig:faultClassLow_multi} summarizes the outcomes across High, Medium, and Low probabilities.

Regarding registers and L1 caches, as expected, the "Crash" rate becomes the overwhelming outcome for \textit{CHAOSReg}, \textit{CHAOSCacheL1I}, and \textit{CHAOSCacheL1D}, even at lower injection probabilities compared to the single-bit scenario. The probability of flipping multiple bits within a valid instruction word or a pointer address almost invariably results in an illegal state or segmentation fault. \textit{Bitcount} and \textit{Qsort} remain the most fragile workloads, exhibiting crash rates near 100\% in High probability scenarios.

Remarkable resilience is observed in main memory. Even with multi-bit corruption, the crash rate remains negligible across all benchmarks. The dominant outcomes remain Masked and SDC. This suggests that the vast address space of the main memory effectively "dilutes" the impact of multi-bit faults; corrupting multiple bits in an unused data array or padding area is statistically more probable than hitting a critical control structure.

\subsubsection{HPC Variation and System Stability}

The analysis of HPCs reveals that multi-bit faults induce more extreme performance anomalies than single-bit faults, particularly in sorting algorithms. The results are detailed in Tables~\ref{tab:chaosreg_sdc_masked_multi} through \ref{tab:chaosmem_sdc_masked_multi}.

\begin{table*}[]
    \centering
    \small
    
    \caption{Analysis of HPC Variations under Multi-Bit Injection: \textit{CHAOSReg}}
    \label{tab:chaosreg_sdc_masked_multi}
    \resizebox{\linewidth}{!}{
    \begin{tabular}{@{}ccccccccccccccccc@{}}
    \toprule
    \textbf{} & \multicolumn{2}{c}{\textbf{Bitcount}} & \multicolumn{2}{c}{\textbf{Blowfish}} & \multicolumn{2}{c}{\textbf{Dijkstra}} & \multicolumn{2}{c}{\textbf{JPEG}} & \multicolumn{2}{c}{\textbf{Patricia}} & \multicolumn{2}{c}{\textbf{Qsort}} & \multicolumn{2}{c}{\textbf{SHA}} & \multicolumn{2}{c}{\textbf{Susan}} \\
    \midrule
     & SDC & Masked & SDC & Masked & SDC & Masked & SDC & Masked & SDC & Masked & SDC & Masked & SDC & Masked & SDC & Masked \\
    \textbf{High} & 6.77 & - & - & - & 6.65 & - & 18.28 & - & 0.00 & - & 0.00 & - & - & - & - & - \\
    \textbf{Medium} & 1.09 & 0.00 & 0.83 & 0.86 & 4.43 & 0.00 & 17.99 & 0.02 & 0.04 & 0.00 & 0.00 & 0.00 & 0.00 & 0.00 & 0.15 & - \\
    \textbf{Low} & 2.11 & 0.00 & 0.93 & 0.86 & 8.23 & 0.00 & 0.02 & 0.02 & 0.01 & 0.00 & 42912.66 & 0.00 & 0.00 & 0.00 & 0.44 & - \\
    \bottomrule
    \end{tabular}
    }
    
    \vspace{0.2cm}
    
    \caption{Analysis of HPC Variations under Multi-Bit Injection: \textit{CHAOSCacheL1D}}
    \label{tab:chaoscachel1d_sdc_masked_multi}
    \resizebox{\linewidth}{!}{
    \begin{tabular}{@{}ccccccccccccccccc@{}}
    \toprule
    \textbf{} & \multicolumn{2}{c}{\textbf{Bitcount}} & \multicolumn{2}{c}{\textbf{Blowfish}} & \multicolumn{2}{c}{\textbf{Dijkstra}} & \multicolumn{2}{c}{\textbf{JPEG}} & \multicolumn{2}{c}{\textbf{Patricia}} & \multicolumn{2}{c}{\textbf{Qsort}} & \multicolumn{2}{c}{\textbf{SHA}} & \multicolumn{2}{c}{\textbf{Susan}} \\
    \midrule
     & SDC & Masked & SDC & Masked & SDC & Masked & SDC & Masked & SDC & Masked & SDC & Masked & SDC & Masked & SDC & Masked \\
    \textbf{High} & 0.02 & - & 0.63 & - & 0.01 & - & 0.02 & - & 0.01 & - & 15732.26 & 0.00 & 0.09 & - & 0.08 & - \\
    \textbf{Medium} & 1.37 & 0.01 & 1.43 & - & 0.01 & 0.00 & 0.04 & 0.02 & 0.01 & 0.00 & 81920.37 & 0.00 & 0.01 & - & 0.07 & - \\
    \textbf{Low} & 0.03 & 0.02 & 0.78 & 0.83 & 0.10 & 0.00 & 0.10 & 0.02 & 2.70 & 0.00 & 81817.61 & 0.00 & 0.02 & 0.00 & 0.10 & - \\
    \bottomrule
    \end{tabular}
    }
    
    \vspace{0.2cm}
    
    \caption{Analysis of HPC Variations under Multi-Bit Injection: \textit{CHAOSCacheL1I}}
    \label{tab:chaoscachel1i_sdc_masked_multi}
    \resizebox{\linewidth}{!}{
    \begin{tabular}{@{}ccccccccccccccccc@{}}
    \toprule
    \textbf{} & \multicolumn{2}{c}{\textbf{Bitcount}} & \multicolumn{2}{c}{\textbf{Blowfish}} & \multicolumn{2}{c}{\textbf{Dijkstra}} & \multicolumn{2}{c}{\textbf{JPEG}} & \multicolumn{2}{c}{\textbf{Patricia}} & \multicolumn{2}{c}{\textbf{Qsort}} & \multicolumn{2}{c}{\textbf{SHA}} & \multicolumn{2}{c}{\textbf{Susan}} \\
    \midrule
     & SDC & Masked & SDC & Masked & SDC & Masked & SDC & Masked & SDC & Masked & SDC & Masked & SDC & Masked & SDC & Masked \\
    \textbf{High} & 3847.78 & - & 0.78 & - & 0.00 & - & 7.90 & - & 0.01 & - & 20.09 & - & 28.99 & - & 4.78 & - \\
    \textbf{Medium} & 0.04 & - & 0.53 & - & 2.33 & - & 5.23 & 0.02 & 1.05 & 0.00 & 0.81 & 0.00 & 0.12 & - & 0.66 & - \\
    \textbf{Low} & 12.38 & - & 0.72 & 0.83 & 0.06 & 0.00 & 0.04 & 0.02 & 9.82 & 0.00 & 1.13 & 0.00 & 28.96 & 0.00 & 0.02 & - \\
    \bottomrule
    \end{tabular}
    }
    
    \vspace{0.2cm}
    
    \caption{Analysis of HPC Variations under Multi-Bit Injection: \textit{CHAOSCacheL2}}
    \label{tab:chaoscachel2_sdc_masked_multi}
    \resizebox{\linewidth}{!}{
    \begin{tabular}{@{}ccccccccccccccccc@{}}
    \toprule
    \textbf{} & \multicolumn{2}{c}{\textbf{Bitcount}} & \multicolumn{2}{c}{\textbf{Blowfish}} & \multicolumn{2}{c}{\textbf{Dijkstra}} & \multicolumn{2}{c}{\textbf{JPEG}} & \multicolumn{2}{c}{\textbf{Patricia}} & \multicolumn{2}{c}{\textbf{Qsort}} & \multicolumn{2}{c}{\textbf{SHA}} & \multicolumn{2}{c}{\textbf{Susan}} \\
    \midrule
     & SDC & Masked & SDC & Masked & SDC & Masked & SDC & Masked & SDC & Masked & SDC & Masked & SDC & Masked & SDC & Masked \\
    \textbf{High} & 47.82 & 0.00 & 96.34 & 0.68 & 0.02 & 0.00 & 0.58 & 0.02 & 5.19 & 0.00 & 23822.25 & 0.00 & 0.03 & - & 0.03 & - \\
    \textbf{Medium} & 0.71 & 0.00 & 0.75 & 0.75 & 1.40 & 0.00 & 0.08 & 0.02 & 7.13 & 0.00 & 392.66 & 0.00 & 0.04 & 0.00 & 0.01 & - \\
    \textbf{Low} & 2.20 & 0.12 & 0.72 & 0.72 & 0.52 & 0.00 & 0.02 & 0.02 & 0.08 & 0.04 & 1819.95 & 0.00 & 0.04 & 0.00 & 0.01 & - \\
    \bottomrule
    \end{tabular}
    }
    
    \vspace{0.2cm}
    
    \caption{Analysis of HPC Variations under Multi-Bit Injection: \textit{CHAOSMem}}
    \label{tab:chaosmem_sdc_masked_multi}
    \resizebox{\linewidth}{!}{
    \begin{tabular}{@{}ccccccccccccccccc@{}}
    \toprule
    \textbf{} & \multicolumn{2}{c}{\textbf{Bitcount}} & \multicolumn{2}{c}{\textbf{Blowfish}} & \multicolumn{2}{c}{\textbf{Dijkstra}} & \multicolumn{2}{c}{\textbf{JPEG}} & \multicolumn{2}{c}{\textbf{Patricia}} & \multicolumn{2}{c}{\textbf{Qsort}} & \multicolumn{2}{c}{\textbf{SHA}} & \multicolumn{2}{c}{\textbf{Susan}} \\
    \midrule
     & SDC & Masked & SDC & Masked & SDC & Masked & SDC & Masked & SDC & Masked & SDC & Masked & SDC & Masked & SDC & Masked \\
    \textbf{High} & - & 0.00 & - & 0.86 & 2.05 & 0.00 & 0.02 & 0.02 & 0.00 & 0.00 & 0.60 & 0.00 & 0.02 & 0.00 & 0.01 & - \\
    \textbf{Medium} & - & 0.00 & - & 0.86 & - & 0.00 & 0.02 & 0.02 & - & 0.00 & 0.00 & 0.00 & - & 0.00 & 0.01 & - \\
    \textbf{Low} & - & 0.00 & - & 0.86 & 0.00 & 0.00 & 0.02 & 0.02 & - & 0.00 & 0.00 & 0.00 & - & 0.00 & 0.01 & - \\
    \bottomrule
    \end{tabular}
    }
\end{table*}

\begin{itemize}
    \item The most striking result is observed in \textit{Qsort}. Under \textit{CHAOSReg} injection at Low probability, the HPC variation spikes to an extraordinary 42,912\%. This is a distinct departure from the single-bit case, where register faults caused only minor variances. This suggests that multi-bit corruption in registers likely destroys loop boundaries or array indices, forcing the sorting algorithm into a pathological state that executes millions of extra instructions without crashing. Similarly, L1D Cache faults for \textit{Qsort} maintain extreme variations, exceeding 81,000\% in Medium and Low scenarios.

    \item \textit{CHAOSCacheL2} shows increased sensitivity. At High probabilities, \textit{Qsort} exhibits a variation of 23,822\%, and \textit{Blowfish} reaches nearly 96\%, indicating that aggressive corruption in the secondary cache propagates to the core execution pipeline more effectively than single-bit errors.

    \item Consistent with the outcome distribution, \textit{CHAOSMem} shows negligible HPC variations (mostly $<0.02\%$) across all probabilities. This confirms that memory faults are "silent" not only in terms of crashes but also in micro-architectural activity. They corrupt data values without perturbing the execution flow, making them the hardest to detect via performance counters.
\end{itemize}

Random multi-bit injection polarizes the system behavior: it either kills the application immediately (registers/L1) or hides silently in the memory background, with specific algorithmic outliers like \textit{Qsort} demonstrating that "surviving" a fault can sometimes be more computationally expensive than crashing.

\section{Conclusions}
\label{sec:conclusions}

\textit{CHAOS} provides a practical and flexible framework for fault injection in the evaluation of system reliability and resilience.

In a landscape where existing tools rapidly become obsolete, often losing compatibility with current gem5 versions and suffering from functional limitations, \textit{CHAOS} emerges as a robust, open-source framework that is inherently configurable and extensible.

Our investigation reveals the nuanced mechanisms of system vulnerability. The experimental results demonstrate that fault injection generates far more complex consequences than traditional binary failure models. Silent faults can induce substantial variations in hardware performance counters, often without triggering obvious system failures, a critical observation for understanding system resilience.

\textit{CHAOS}'s modular architecture enables comprehensive exploration of fault scenarios across multiple architectural levels. Diverse case studies, including implementations like JPEG and Qsort, expose remarkable variations in system robustness, with some fault injection probabilities even revealing the potential for complete system breakdown.

These observations are increasingly relevant given the growing complexity of computational systems used in critical domains. Beyond serving as a research tool, \textit{CHAOS} offers a structured methodology to support the analysis and improvement of system-level fault tolerance.

Future research directions include expanding the taxonomy of simulated faults, developing more advanced error detection techniques, and exploring integration with emerging computational architectures.

\section*{Acknowledgment}
This work has been (partially) supported by the Spoke 1 "FutureHPC \& BigData" of the Italian Research Center on High-Performance Computing, Big Data and Quantum Computing (ICSC) funded by MUR Missione 4 - Next Generation EU (NGEU) and MUR PRIN COLTRANE-V E53D23008060006.

\bibliographystyle{IEEEtran}
\bibliography{bibliography}

\end{document}